\documentclass[aps,prd,twoside,twocolumn,floatfix,letterpaper]{revtex4-1}
\usepackage[portuguese]{babel}
\usepackage[utf8x]{inputenc}
\usepackage[dvipdfm]{graphicx}
\usepackage[dvipdfm]{rotating}
\usepackage{graphicx,pdfpages}
\usepackage{subfig}
\usepackage{float}
\usepackage{lipsum,fancyhdr}
\usepackage{textcomp}
\usepackage{subfig}
\usepackage{array}
\usepackage{amssymb, amsmath, pxfonts}
\usepackage[T1]{fontenc}
\usepackage{mathptmx}
\usepackage{mathtools}

\fancypagestyle{plain}{%
\chead{}

}

\newcommand{\hcond}{\usefont{T1}{phv}{mc}{n}}
\makeatletter
\def\section{\@startsection{section}{1}{0pt}{-3.5ex plus -1ex minus -.2ex}{2.3ex plus .2ex}{\large \hcond}}
\def\subsection{\@startsection{subsection}{2}{\z@}{-3.25ex plus -1ex minus -.2ex}{1.5ex plus .2ex}{\hcond}}

\begin{document}

\title{Equivalência entre Mecânica Quântica e a Mecânica Quântica PT Simétrica.\\
\footnotesize{Equivalence between Quantum Mechanics and PT Symmetric Quantum Mechanics.}}

\author{David Girardelli, Eduardo M. Zavanin, Marcelo M. Guzzo} \email[]{davidgi@ifi.unicamp.br \\ zavanin@ifi.unicamp.br \\ guzzo@ifi.unicamp.br}
\affiliation{Instituto de F\'\i sica
  Gleb Wataghin \\ Universidade Estadual de Campinas - UNICAMP \\
  Rua S\'ergio Buarque de Holanda, 777 \\
  13083-859 Campinas SP Brazil}

\setcounter{page}{1}

\begin{abstract}

Este artigo traz uma discussão a respeito da Mecânica Quântica PT Simétrica, desenvolvendo  alguns elementos básicos dessa teoria. Em um caso simples de sistema de dois níveis, desenvolvemos o problema da Braquistócrona Quântica. Comparando os resultados obtidos entre a Mecânica Quântica PT Simétrica aqueles obtidos usando o formalismo padrão, concluí-se que a nova abordagem não é capaz de revelar nenhum fenômeno novo.\\
\textbf{Palavras-chave:} Simetria PT, Braquistócrona Quântica, Mecânica Quântica Não Hermitiana.
\\
\\
In this paper we develop a discussion about PT Symmetric Quantum Mechanics, working with basics elements of this theory. In a simple case of two body system, we developed the Quantum Brachistochrone problem. Comparing the results obtained through the PT Symmetric Quantum Mechanics with that ones obtained using the standard formalism, we conclude that this new approach is not able to reveal any new effect.
\end{abstract}

\maketitle\thispagestyle{plain}

\section{Introdução}

É comum em Mecânica Quântica definirmos a Hamiltoniana que descreve a evolução temporal dos estados
de maneira hermitiana ($H = H^{\dag}$). Esta imposição garante que os autovalores correspondentes a H e, portanto, os níveis de energia dos sistema, sejam reais. Tal imposição garante a evolução unitária do sistema que, por sua vez, implica que as probabilidades derivadas dessa hamiltoniana sejam preservadas.

No entanto, apesar da condição de hermiticidade ser uma condição suficiente para se obter autovalores reais, ela não é necessária. É possível construir uma teoria quântica com hamiltonianas que, apesar de não serem hermitianas, possuam autovalores reais, provendo uma evolução unitária para o sistema.

O estudo desse tipo de Hamiltonianas começou em 1998 com Bender e Boetcher \cite{PhysRevLett.80.5243}, que propuseram a substituição da condição de hermiticidade por uma condição análoga, chamada de Simetria PT, $PT H (PT)^{-1} = H$, (neste contexto P se refere ao operador paridade enquanto T se refere ao operador reflexão temporal) que garantia que os níveis de energia do sistema se mantivessem reais. Após a publicação deste artigo seminal, outros relacionados a este tema foram desenvolvidos, estudando uma série de diferentes hamiltonianas com o intuito de encontrar peculiaridades na Mecânica Quântica PT Simétrica, que possivelmente explicariam fenômenos físicos ainda não entendidos \cite{PhysRevLett.80.5243,PhysRevLett.93.251601,Znojil2001mc,PhysRevLett.89.270401,1402-4896-87-1-017001,doi1451022,doi1689673}. Em \cite{mostafazadehgeneral}, Mostafazadeh generalizou o conceito de Simetria PT, encontrando relações gerais nas quais matrizes não hermitianas possuíssem autovalores reais,  definindo isso como Pseudo Hermiticidade. 
No presente artigo, vamos desenvolver elementos da Mecânica Quântica PT Simétrica, obtendo operadores básicos para essa teoria. Além disso,  vamos analisar a Braquistócrona Quântica sob o ponto de vista da Mecânica Quântica Usual e da PT Simétrica  \cite{PhysRevLett.96.060503,onoptimun,faster}. Através dessa análise, podemos ser induzidos a acreditar em comportamentos anômalos, que geram transições infinitamente rápidas entre estados quânticos \cite{faster}, vamos entender esse fenômeno e identificar que essa transição rápida se dá devido a uma interpretação equivocada dos estados quânticos.

\section{Mecânica Quântica Usual}

A Mecânica Quântica pressupõe alguns pontos importante que vamos recordar. Em seguida, iremos compará-los com aqueles da Mecânica Quântica PT Simétrica. 

\begin{enumerate}

\item \textit{Autofunções e Autovalores}

As energias e os estados acessíveis do sistema, são obtidos através das soluções da Equação de Schrödinger:
\begin{equation}
\label{sch}
 H|\psi\rangle = i\hbar \frac{\partial |\psi \rangle}{\partial t}, 
 \end{equation}
Para o caso particular onde a Hamiltonia H não seja dependente do tempo, podemos resolver a Eq. (\ref{sch}), obtendo:
 \begin{equation}
  |\psi (t) \rangle = e^{\frac{-i}{\hbar}\int H dt}|\psi\rangle, \\
 \end{equation}
 \begin{equation}
  |\psi (t) \rangle = U|\psi\rangle \label{U},
 \end{equation}

 onde $U$ é denominado operador de evolução temporal, definido por $U \equiv e^{\frac{-i}{\hbar}\int H dt}$.

\item \textit{Completeza}

A Completeza oferece uma maneira de representar qualquer estado como uma combinação linear de autoestados que formam uma base para o espaço:

\begin{eqnarray}
\sum_{n=0}^{\infty}[\psi_n(y)]^{\dagger} \psi_n(x) =\delta(x-y), \\
\sum_{n=0}^{\infty}[\langle  y | \psi_n \rangle_D]^{\dagger}  \langle x | \psi_n \rangle_D = \delta(x-y), \\
\sum_{n=0}^{\infty}\langle \psi_n  | y \rangle_D  \langle x | \psi_n \rangle_D = \langle x |I| y \rangle_D. 
\label{troca}
\end{eqnarray}

Onde utilizamos a notação $\psi_n(x) = \langle x | \psi_n \rangle_D$. Note que na Eq. (\ref{troca}), $\langle \psi_n  | x \rangle_D$ e $\langle y  | \psi_n \rangle_D$ são escalares e, portanto, podem trocar de lugar dentro do somatório:

\begin{eqnarray}
\label{troca2}
\sum_{n=0}^{\infty}\langle x | \psi_n \rangle_D  \langle \psi_n  | y \rangle_D  = \langle x |I| y \rangle_D, \\
\sum_{n=0}^{\infty}| \psi_n \rangle  \langle \psi_n| = I,
\label{completeza}
\end{eqnarray}

ou, multiplicando por $|\chi \rangle$ pela direita:

\begin{equation}
\sum_{n=0}^{\infty}| \psi_n \rangle  \langle \psi_n|\chi \rangle_D = |\chi \rangle.
\label{completeza}
\end{equation}

Que deixa claro que um estado genérico $|\chi\rangle$ pode ser representado por uma combinação linear de autoestados $| \psi_n \rangle$, com os devidos coeficientes $\langle \psi_n|\chi\rangle_D$.

\item \textit{Produto Interno e Ortonormalidade das Autofunções}

O produto interno para Mecânica Quântica Usual é definido por:

\begin{equation}
\label{orto}
\langle \psi| \psi \rangle _D \equiv \int{ [\psi(x)]^{\dagger} \psi(x) dx}.
\end{equation}

Neste artigo o índice D será sempre designado a operações utilizando o produto interno de Dirac comum na Mecânica Quântica Padrão.

A ortonormalidade é obtida quando todos autoestados quânticos da Hamiltoniana possuem norma unitária e o produto interno entre dois autoestados distintos é nulo,

\begin{equation}
\label{deltadirac}
\langle \psi_m|\psi_n\rangle_D = \delta_{mn}.
\end{equation}

\item \textit{Unitariedade}

Seja um estado genérico $|\chi\rangle$. Utilizando a hermiticidade de $H$ ($H^{\dag} = H$), e que a evolução temporal de um sistema quântico é descrita pela Eq. (\ref{U}), podemos calcular a 
norma da evolução temporal de $|\chi \rangle$,

\begin{equation}
\langle \chi(t)|\chi(t)\rangle_D = \langle \chi| U^{\dagger}U |\chi \rangle_D = \langle \chi | \chi\rangle_D = 1 .
\label{unit}
\end{equation}

A Eq. (\ref{unit}) nos mostra que, independentemente da evolução temporal de um estado quântico, a norma deste é sempre preservada.
\end{enumerate}

Da mesma forma que em Mecânica Quântica Usual, enunciaremos pontos importantes da Mecânica Quântica PT Simétrica.

\section{Mecânica Quântica PT Simétrica}
\begin{enumerate}

\item \textit{Autofunções e Autovetores}

Os autovalores e autovetores obtidos para Mecânica Quântica PT Simétrica também são provenientes da solução da equação de Schröedinger onde, nesse caso, o operador H na Eq. (\ref{sch}) é composto por uma Hamiltoniana PT Simétrica. No que diz respeito a evolução temporal de um sistema quântico, na Simetria PT, também é postulada a equação de Schrodinger, descrita pela Eq. (\ref{sch}).

\item \textit{Completeza}

A relação de completeza para a Mecânica Quântica PT Simétrica se apresenta da seguinte maneira:

\begin{eqnarray}
 \sum_{n=0}^{\infty}(-1)^n\psi_n(x)\psi_n(y) = \delta(x-y).
 \label{completezanh}
 \end{eqnarray}

Essa relação nada intuitiva foi testada numericamente e analiticamente \cite{0305-4470-33-27-308,0305-4470-34-15-401} para uma certa classe de teorias desenvolvidas em \cite{makingsense}, todavia, uma prova matemática da mesma pode ser encontrada em \cite{PhysRevA.68.062111}.

\item \textit{Produto Interno e Ortonormalidade das Autofunções}

O produto interno no caso Hermitiano é descrito pela Eq. (\ref{orto}). Para o caso da Simetria PT, a definição de produto interno não é igual ao caso padrão. Uma possível abordagem para construir o produto interno seria partir da transformação $H^{PT}= H$, sendo que $H^{PT}$ designa a operação $H^{PT}=PT (H) (PT)^{-1}$.
Similarmente ao produto interno de Dirac, o produto interno não Hermitiano consiste em substituir a operação transposto conjugado ($\dagger$), pelo operador paridade-reversão temporal ($PT$), desta maneira definimos o produto interno $PT$ como: 

\begin{eqnarray}
\label{pt}
\langle \psi_n | \psi_m \rangle_{PT} =  \int [\psi_n(x)]^{PT}\psi_m(x) dx, \\
\langle \psi_n | \psi_m \rangle_{PT} =  \int [PT\psi_n(x)]\psi_m(x)dx,  \\
\langle \psi_n | \psi_m \rangle_{PT} =  \int \psi_n(-x)^{*}\psi_m(x)dx.
\label{comp}
\end{eqnarray}

Uma vez que $\psi_n(x)$ e $\psi_m(x)$ são autoestados e soluções independentes de (\ref{sch}) com autovalores distintos, o produto interno entre elas é nulo sempre que $n \neq m$. 

Diferentemente do esperado, a norma dos estados $\psi_n(x)$s é ligeiramente diferente, e é definido da seguinte maneira:

\begin{eqnarray}
\langle \psi_n | \psi_n \rangle_{PT} =  \int \psi_n(-x)^{*}\psi_n(x)dx = (-1)^n.
\label{pt2}
\end{eqnarray}

Essa relação também é verificada numericamente em \cite{0305-4470-33-27-308,0305-4470-34-15-401}.

Como conclusão imediata vemos que há estados que possuem norma negativa ($n$ ímpares) e também há estados que possuem norma positiva ($n$ pares). 

Para sanar o problema referente às normas negativas é necessário a introdução de um novo operador. Este operador terá matematicamente o objetivo de incrementar um fator na Eq. (\ref{pt2}) que cancele o termo $(-1)^n$. O problema aqui apresentado é muito similar ao que Dirac se deparou ao formular a equação de onda espinorial em Teoria Quântica Relativística onde, para cada estado com energia positiva, havia um outro estado com energia negativa. Este problema foi resolvido postulando a existência de antipartículas, de forma que partículas e antipartículas se relacionavam através de um operador, denominado conjugação de carga (C). Analogamente, para toda Hamiltoniana PT Simétrica há uma simetria que relaciona cada estado de norma negativa a outro com norma positiva e, denominaremos este operador como $\mathcal{C}$. Note que C e $\mathcal{C}$ não possuem o mesmo significado. C conecta antipartículas com partículas no contexto de Teoria Quântica de Campos enquanto que $\mathcal{C}$ conecta estados de normas negativas com estados de normas positivas, no contexto de Mecânica Quântica PT Simétrica, onde estamos lidando com a equação de Schrödinger, que descreve a função de onda de apenas uma partícula. Em Mecânica Quântica PT Simétrica, $\mathcal{C}$ pode ser definido como soma sobre todas as autofunções normalizadas da Hamiltoniana,

\begin{equation}
 \mathcal{C}(x,y) =  \sum_{n=0}^\infty \psi_n(x)\psi_n(y). \\
\label{maisumac}
\end{equation}

Para entender como o operador $\mathcal{C}$ incrementa o fator $(-1)^n$, podemos aplicá-lo a uma autofunção $\psi_n(x)$:

\begin{equation}
\langle x| \mathcal{C} |\psi_n\rangle = \int \langle x |\mathcal{C}|y\rangle \langle y| \psi_n \rangle dy = \int \mathcal{C}(x,y) \psi_n (y) dy,
\label{cdefin2}
\end{equation}
onde, na última passagem utilizamos $\int |y\rangle\langle y|dy = I$, uma vez que $|y\rangle$ são autofunções do operador posição $y$, que é hermitiano, as relações de completeza são exatamente iguais àquelas utilizadas em Mecânica Quântica Usual.
Utilizando a Eq. (\ref{cdefin2}) juntamente com Eq. (\ref{maisumac}):

\begin{eqnarray}
\langle x| \mathcal{C} |\psi_n\rangle = \sum_{m=0}^\infty \int \psi_m(x)\psi_m(y)\psi_n(y)dy, \\
\langle x| \mathcal{C} |\psi_n\rangle = \sum_{m=0}^\infty \int \frac{(-1)^m}{(-1)^m}\psi_m(x)\psi_m(y)\psi_n(y)dy,
  \end{eqnarray}

usando a relação de completeza (\ref{completezanh}), para as funções $\psi_m$ :

\begin{eqnarray}
\langle x| \mathcal{C} |\psi_n\rangle = \int \frac{1}{(-1)^m}\delta(x-y)\delta_{m,n}\psi_m(y)dy, \\
\langle x| \mathcal{C} |\psi_n\rangle = (-1)^n\psi_n(x), \\
 \mathcal{C} |\psi_n\rangle = (-1)^n |\psi_n \rangle.
\label{p1}
\end{eqnarray}

Definido o operador $\mathcal{C}$, podemos criar um produto interno, da mesma maneira que criamos o produto interno PT, mas agora utilizando a condição: $H^{\mathcal{C}PT} = H$,

\begin{eqnarray}
\langle \psi_n | \psi_n \rangle_{\mathcal{C}PT} = \int [\psi_n(x)]^{\mathcal{C}PT}\psi_n(x) dx, \\
\langle \psi_n | \psi_n \rangle_{\mathcal{C}PT} = \int [\mathcal{C}PT\psi_n(x)]\psi_n (x) dx .
\label{Cpt}
\end{eqnarray}

Onde,

\begin{eqnarray}
\langle \psi_n | \psi_m \rangle_{\mathcal{C}PT} =  \delta_{nm}.
\label{pt2}
\end{eqnarray}

E, portanto, o produto interno entre os estados é positivo definido, desaparecendo o problema de normas negativas.

\item \textit{Unitariedade}

Para verificar a unitariedade notamos que:
\begin{equation}
H^{PT} = PT H (PT)^{-1} = H,
\label{newhermite}
\end{equation}
e, usando a equação de Schrödinger:
\begin{equation}
i\frac{d}{dt}|\psi(t)\rangle = H |\psi(t)\rangle,
\label{schrevolution}
\end{equation}
aplicando $PT$ pela esquerda e utilizando que $(PT)^{-1} PT = I$, onde $I$ é o operador identidade, obtemos:
\begin{equation}
-i\frac{d}{dt}PT|\psi(t)\rangle = PT H (PT)^{-1} PT |\psi(t)\rangle,
\end{equation}
utilizando a Eq. (\ref{newhermite}):
\begin{equation}
-i\frac{d}{dt}PT|\psi(t)\rangle = H^{PT} PT |\psi(t)\rangle,
\label{evpt}
\end{equation}
e, usando que para o caso de PT Simetria $\langle\psi(t)| = (PT |\psi(t)>)^t$, onde t é referente a operação transpor usual de cálculo matricial, transpondo dos dois lados de (\ref{evpt}), obtemos:
\begin{equation}
-i\frac{d}{dt}\langle\psi(t)| =  \langle\psi(t)|H^{PT}.
\label{ptevolution}
\end{equation}
Para estudar a unitariedade, queremos analisar a evolução da norma $\langle \psi(t)|\psi(t)\rangle_{PT}$ e, portanto:
\begin{eqnarray}
\frac{d}{dt}\langle\psi(t)|\psi(t)\rangle_{PT} = \langle\psi(t)|\frac{d}{dt}|\psi(t)\rangle_{PT} + \\ \nonumber
\left[\frac{d}{dt}\langle\psi(t)\right]|\psi(t)\rangle_{PT},
\end{eqnarray}
utilizando as Eqs. (\ref{ptevolution}) e (\ref{schrevolution}), obtemos:
\begin{eqnarray}
\frac{d}{dt}\langle\psi(t)|\psi(t)\rangle_{PT} = -i [\langle\psi(t)|H|\psi(t)\rangle_{PT} - \\ \nonumber
\langle\psi(t)|H^{PT}|\psi(t)\rangle_{PT} ] = 0,
\end{eqnarray}
e, portanto, vemos que a norma dos vetores não muda com a evolução temporal, o que nos garante a unitariedade.

Após implementados estes conceitos, iremos agora aplicar estas ideias a uma Hamiltoniana não Hermitiana 2x2.

\end{enumerate}

\section{Hamiltoniana PT Simétrica}

\subsection{Autoestados}

Para entendermos melhor a Mecânica Quântica PT Simétrica, vamos desenvolver os elementos básicos dessa teoria para o caso do sistema de dois níveis.

Seja $M$ a matriz quadrada de ordem 2:

\begin{equation}
 M = \left[\begin{array}{c c}
\alpha &  \beta \\
\gamma & \theta \end{array} \right],
\end{equation}

com $\alpha$, $\beta$, $\gamma$ e $\theta$ $\in$ $\mathbb{C}$. Calculando os autovalores $\lambda$ 
desta matriz:

\begin{equation} det M = det \left[\begin{array}{c c}
\alpha - \lambda &  \beta \\
\gamma & \theta - \lambda \end{array} \right] = 0 .
\end{equation}

\begin{eqnarray}
 (\alpha - \lambda)(\theta - \lambda) - \gamma \beta = 0, \nonumber \\
 \lambda^2 -\lambda(\alpha+\theta)-\gamma \beta = 0, \nonumber \\
\lambda = \frac{(\alpha+ \theta) \pm \sqrt{(\alpha+ \theta)^2 + 4 \gamma \beta}}{2} .
\label{autovalorl}
\end{eqnarray}

Como queremos sentido físico para essa Hamiltonina, precisamos que seus autovalores $\lambda$ sejam reais.
Para isso, devemos respeitar as seguintes condições:

\begin{eqnarray}
\label{apluso}
 (\alpha + \theta) \in \mathbb{R}, \\
 (\alpha + \theta)^2 + 4 \gamma \beta \in \mathbb{R^+}.
\label{gammabeta}
\end{eqnarray}

Para garantir a relação (\ref{apluso}), definimos que $\alpha = a + i.b$ com $a$ e $b$ $\in \mathbb{R}$ e 
$\theta = c +i.d$ com $c$ e $d$ $\in \mathbb{R}$, deste modo: $d =-b$.

Para garantir a condição dada pela Eq. (\ref{gammabeta}) é necessário que o produto $\gamma\beta \in \mathbb{R^+}$.
Um das maneiras de satisfazer todas as condições impostas é supor que : 
$\gamma = \beta = s$, $\alpha = re^{i\psi}$ e $\theta = re^{-i\psi}$ com $s$, $r$ e $\psi$ $\in \mathbb{R}$. Desta maneira $Im \alpha = - Im \theta$.

Substituindo em $M$ a parametrização encontrada, obtemos a matriz não Hermitiana $H_{NH}$:

\begin{equation}  H_{NH} = \left[\begin{array}{c c}
r.e^{i\psi} &  s \\
s &  r.e^{-i\psi} \end{array} \right] .
\label{hnm}
\end{equation}

Por construção, esta matriz possui autovalores reais, $\epsilon_+$ e $\epsilon_-$ que são dados por:

\begin{eqnarray}
\label{eps}
\epsilon_{\pm} = rcos\psi \pm \sqrt{s^2 - r^2sen^2\psi} . 
\end{eqnarray}

Note que é explicitamente necessário que para $\epsilon$ ser real, $r$ seja menor ou igual a $s$.
Isto permite definir um ângulo $\alpha$ que satisfaz $(r/s)sen\psi = sen\alpha$. Reescrevendo os autovalores em termos desse novo parâmetro, obtemos:

\begin{eqnarray} 
\label{ep1}
 \epsilon_+ = rcos\psi + scos\alpha, \\
 \epsilon_- = rcos\psi - scos\alpha. 
\label{ep2}
 \end{eqnarray}

Os autovetores não normalizados obtidos por estes autovalores são dados por:

\begin{equation} \overrightarrow{\epsilon_+}  = \left[\begin{array}{c}
e^{\frac{i\alpha}{2}}  \\
e^{\frac{-i \alpha}{2}} \end{array} \right].
\end{equation}

\begin{equation} \overrightarrow{\epsilon_-} = \left[\begin{array}{c}
ie^{\frac{-i\alpha}{2}}  \\
-ie^{\frac{i \alpha}{2}} \end{array} \right].
\end{equation}

Note que, se usarmos a definição de produto interno de Dirac, estes vetores não são ortogonais, mesmo sendo autoestados da Hamiltoniana:

\begin{eqnarray}
\label{naoort1}
 \langle \epsilon_{\pm}|\epsilon_{\mp} \rangle_D \neq 0, \\ 
\langle \epsilon_{\pm}|\epsilon_{\pm} \rangle_D \neq 1.
\label{naoort2}
\end{eqnarray}
 
Assim, se faz necessário redefinir este produto através da imposição das matrizes $\mathcal{C}PT$.
 
 \subsection{Operadores}
 
Redefinir o produto interno com respeito a Simetria PT, pode trazer algumas complicações, principalmente devido a existência de arbitrariedades na escolha das matrizes envolvidas. Uma boa referência para este tipo de problema pode ser encontrada em \cite{Bender20092670}. No entanto, nesse artigo, vamos desenvolver uma abordagem bem simples afim de se obter os operadores para o caso de matrizes $2x2$.

As Matrizes $\mathcal{C}$, $P$, $T$, necessárias para a construção do produto interno devem satisfazer as condições que serão enumeradas a seguir: 

\begin{enumerate}

\item \textit{Operador $T$: Reversão Temporal}

O operador T é conhecido na Mecânica Quântica usual, especialmente em Teoria de Campos. 
A função do operador T é trocar o sinal dos elementos que possuam o número imaginário $i$ em sua estrutura.

\begin{equation}
 T|\phi \rangle = |\phi*\rangle.
\end{equation}

Assim o operador T é o responsável pela transformação $i \rightarrow -i$ em objetos que o sucedem.
Uma maneira intuitiva de entender o que ocorre é lembrar do operador de evolução temporal $U$ descrito por (\ref{unit}).
Aplicar uma transformação de reversão temporal ($t \rightarrow -t$) neste operador é equivalente a apenas conjugá-lo:

\begin{equation}
\label{timerev}
 T U = T\left[exp\left(\frac{-iHt}{\hbar}\right)\right] = exp\left(\frac{i H^* t}{\hbar}\right) = U^{*}.
 \end{equation}

 Devido a isto, o operador T também é comumente denominado operador de conjugação complexa.

\item \textit{Operador Paridade $P$: Reflexão Espacial}

O Operador $P$ é um operador gerador de reflexões espaciais, dadas pela transformação: $\vec{x} \rightarrow -\vec{x}$, onde $\vec{x}$, é o vetor posição.

Usando a relação de completeza descrita por (\ref{completezanh}) e, relembrando que
$\langle x|P|y \rangle = \langle x|-y \rangle = \delta(x+y)$, podemos escrever o operador paridade para o caso de Simetria PT da seguinte maneira:

\begin{equation}
P = \sum_{n=0}^{\infty}{(-1)^n\psi_n(x) \psi_n(-y) = \delta(x+y)},
\end{equation}

que pode ser escrito em sua representação matricial, usando a completeza (\ref{completezanh}), conforme: 

\begin{equation}
\label{pmatricial}
I = \sum_{n=0}^{\infty}(-1)^n|\psi_n\rangle \langle\psi_n|,
\end{equation}
aplicando P pela direita,

\begin{equation}
\label{pmatricial}
P = \sum_{n=0}^{\infty}(-1)^n|\psi_n\rangle \langle\psi_n|P.
\end{equation}

Lembrando que $\langle \psi_n|$ é o conjugado PT simétrico do estado, ou seja,  $\langle\psi_n| =  (PT|\psi_n>)^{t}$, sendo $t$ o símbolo de transposto e, usando o fato que $(AB)^{t} = B^{t} A^{t}$ e também que $P^t P = I$, podemos escrever:

\begin{eqnarray}
P = \sum_{n=0}^{\infty}(-1)^n|\psi_n\rangle (T|\psi_n\rangle)^{t}, \\
P = (-1)|\epsilon_- \rangle (T|\epsilon_-\rangle)^{t} + |\epsilon_+\rangle(T |\epsilon_+\rangle)^t ,\\
P = (-1)|\epsilon_- \rangle (|\epsilon_-\rangle^*)^{t} + |\epsilon_+\rangle(|\epsilon_+\rangle^*)^t .
\label{pma}
\end{eqnarray}

Note que para obter os operadores $P$ e $\mathcal{C}$ é necessário definirmos os vetores normalizados $|\epsilon_+\rangle$ e $|\epsilon_-\rangle$, deste modo:

\begin{eqnarray}
|\epsilon_+ \rangle = a \vec{\epsilon_+} ,\\
|\epsilon_-\rangle = b \vec{\epsilon_-},
\end{eqnarray}
com $a$ e $b$ sendo constantes reais, sem perda de generalidade.
Substituindo os valores dos autoestados normalizados na Eq. (\ref{pma}), temos como resultado o operador $P$ em função dos parâmetros $a$ e $b$:

\begin{equation}
 P = \left[\begin{array}{c c}
a^2-b^2 & b^2e^{-i\alpha}+a^2e^{i\alpha} , \\
 a^2e^{-i\alpha}+b^2e^{i\alpha} &  a^2-b^2
 \end{array} \right].
\end{equation}

 \item \textit{Operador $\mathcal{C}$}

Para se obter a representação matricial do operador $\mathcal{C}$, utilizamos a Eq. (\ref{maisumac}).

\begin{eqnarray}
\mathcal{C} = \sum_{n}{|\psi_n\rangle \langle\psi_n|}, \\
\mathcal{C} = \sum_{n}{|\psi_n\rangle (PT|\psi_n\rangle)^{t}}, \\
\mathcal{C} = |\epsilon_-\rangle (P|\epsilon_-\rangle^*)^{t}+|\epsilon_+\rangle (P|\epsilon_+\rangle^*)^{t}.
\end{eqnarray}

Usando o mesmo procedimento para o fator de normalização realizado em $P$, segue que:

\begin{equation} \mathcal{C} = 2 \left[\begin{array}{c c}
  a^4-b^4 + ia^2b^2\sin(2\alpha) & a^4e^{i\alpha} +b^4e^{-i\alpha}, \\
   a^4e^{-i\alpha} + b^4e^{i\alpha} &  a^4-b^4 - ia^2b^2\sin(2\alpha)
  \end{array}\right].
  \end{equation}
 \end{enumerate}
 
Com os operadores $P$ e $\mathcal{C}$ em função dos parâmetros $a$ e $b$, finalmente definimos a operação $\mathcal{C}PT$ e, portanto, a construção de nosso produto interno é dado pela Eq. (\ref{Cpt}).

 \section{Normalização $\mathcal{C}$PT}

Sejam  $|\epsilon_+ \rangle$ e $|\epsilon_- \rangle$ os autoestados normalizados dos vetores $\vec{\epsilon_+}$ e $\vec{\epsilon_-}$, respectivamente. 

Como $|\epsilon_+ \rangle$ e $|\epsilon_-\rangle$ são autoestados da Hamiltoniana PT Simétrica, eles são ortonormais com respeito a $\mathcal{C}PT$. Desta maneira, os estados devem satisfazer as condições de ortonormalidade:

 \begin{eqnarray}
 \label{state1}
 \langle \epsilon_{\mp} |\epsilon_{\mp} \rangle_{\mathcal{C}PT} = 1, \\
 \langle \epsilon_{\pm} |\epsilon_{\mp} \rangle_{\mathcal{C}PT} = 0.
 \label{state2}
\end{eqnarray}

Destas condições podemos calcular explicitamente os parâmetros $a$ e $b$.
Aplicando a Eq. (\ref{Cpt}) a condição (\ref{state2}), segue que:

\begin{equation}
 -16a^3(a-b)b^3(a+b)\cos^2 \alpha\sin\alpha = 0 .
 \label{condicao}
\end{equation}

Na Eq. (\ref{condicao}), sabemos que $\alpha$ é um parâmetro livre da Hamiltoniana e pode assumir qualquer valor, além disso, os parâmetros $a$ e $b$, não podem ser nulos, pois assim  $|\epsilon_{\mp} \rangle$ seriam nulos e não formariam uma base ortogonal. Deste modo, as únicas maneiras de se satisfazer a Eq. (\ref{condicao}) são dadas por:

\begin{eqnarray}
(a+b) = 0 \label{condicao1} \Leftrightarrow b=-a, \\
(a-b) = 0 \label{condicao2} \Leftrightarrow b=a.
\end{eqnarray}	

Nota-se que, tanto a escolha da solução (\ref{condicao1}) ou da (\ref{condicao2}), irá gerar o mesmo resultado, pois os operadores em questão dependem apenas de fatores quadráticos e quárticos destes parâmetros. 
Isto sempre mantém uma ambiguidade em relação ao sinal dos parâmetros. Assim, sem perda de generalidade, escolhendo $a=b$, positivos, e aplicando este resultado na Eq. (\ref{state1}), segue que:

\begin{equation}
16a^8\cos^4 \alpha =1,
\end{equation}

logo:

\begin{equation}
 a = \frac{1}{\sqrt{2\cos\alpha}}.
\end{equation}

Deste modo os operadores $P$ e $\mathcal{C}$ normalizados são dados por:

\begin{equation}
 \mathcal{C} = \frac{1}{\cos\alpha}\left[\begin{array}{c c}
i\sin\alpha & 1  \\
1 & -i \sin\alpha \end{array} \right].
\end{equation}

\begin{equation}
P = \left[\begin{array}{c c}
0 & 1  \\
1 & 0 \end{array} \right].
\end{equation}

E, finalmente, os autoestados normalizados são dados por:

\begin{equation} |\epsilon_+ \rangle = \frac{1}{\sqrt{2cos\alpha}}\left[\begin{array}{c}
e^{\frac{i\alpha}{2}}  \\
e^{\frac{-i \alpha}{2}} \end{array} \right],
\end{equation}

\begin{equation} |\epsilon_- \rangle = \frac{1}{\sqrt{2cos\alpha}}\left[\begin{array}{c}
ie^{\frac{-i\alpha}{2}}  \\
-ie^{\frac{-i \alpha}{2}} \end{array} \right].
\end{equation}

Onde os operadores normalizados obedecem as seguintes regras de comutação:

 \begin{eqnarray}
\label{c2}
 \mathcal{C}^2 = I,  \\
\label{ch}
 \left[\mathcal{C},H \right] = 0, \\
\label{cpt}
 \left[\mathcal{C},PT \right] = 0.
\end{eqnarray}

É importante ressaltar que vetores que eram ortogonais com respeito ao produto interno de Dirac, descrito pela Eq. (\ref{orto}), podem não ser ortogonais com respeito a esse novo produto interno,
como por exemplo os estados $|\nu_1 \rangle$ e $|\nu_2 \rangle$,
   
\begin{equation}
|\nu_1 \rangle = \left[\begin{array}{c}
1   \\
0  \end{array} \right], 
\end{equation}        
 
 \begin{equation}
|\nu_2 \rangle = \left[\begin{array}{c}
0   \\
1  \end{array} \right]. 
\end{equation}

Estes estados são evidentemente ortogonais com respeito ao produto interno convencional, dado que: $\langle \nu_1|\nu_2 \rangle_D = 0$. Entretanto, para o produto $\mathcal{C}PT$ temos o seguinte cálculo:

\begin{equation}
\langle \nu_2 |\nu_1 \rangle_{\mathcal{C}PT} =(\mathcal{C}PT|\nu_2\rangle)^t |\nu_1\rangle = - i.tan\alpha . 
\end{equation}

Se $\alpha \neq h.\pi$ com $h$ $\in$ $\mathbb{Z}$, $tan\alpha \neq 0$ e, portanto, os estados $|\nu_1 \rangle$ e $|\nu_2 \rangle$ não são mais ortogonais e dependem explicitamente do parâmetro $\alpha$ da Hamiltoniana.

\section{Evolução Temporal}

A Hamiltoniana estudada por \cite{faster} se apresenta como capaz de obter um comportamento anômalo com respeito a transição de estados, segundo os autores, é possível obter uma transição de maneira mais rápida do que no caso da Mecânica Quântica Usual.

Para entender melhor este fenômeno, vamos detalhar os raciocínios desenvolvidos em \cite{faster}; afim de relacionar a Mecânica Quântica Usual com a PT Simétrica, vamos calcular o tempo de transição entre os estados nas duas situações. O cálculo do tempo mínimo de transição entre estados é o famoso problema da Braquistócrona Quântica.

Usando a equação de Schrödinger para governar a evolução temporal do sistema, que possui como Hamiltoniana a matriz $H_{NH}$, e supondo que os parâmetros $\alpha$, $\beta$, $\gamma$ e $\theta$ não dependam do tempo, a evolução temporal de um estado genérico $|\phi\rangle$ é dada por:

\begin{equation}
 |\phi(t) \rangle = e^{\frac{-iH_{NH}t}{\hbar}}|\phi \rangle,
\label{evolnh}
 \end{equation}
conforme descrito por (\ref{unit}).

Como estamos trabalhando com matrizes $2x2$, podemos expandir a Hamiltoniana em termos das matrizes de Pauli:

\begin{equation}
\sigma_1 = \left[\begin{array}{c c}
0 & 1  \\
1 & 0 \end{array} \right],
\label{sig1}
\end{equation}

\begin{equation}
\sigma_2 = \left[\begin{array}{c c}
0 & -i  \\
i & 0 \end{array} \right],
\label{sig2}
\end{equation}

\begin{equation}
\sigma_3 = \left[\begin{array}{c c}
1 & 0  \\
0 & -1 \end{array} \right],
\label{sig3}
\end{equation}
e da Identidade, afim de calcular o termo exponencial. Assim a Hamiltoniana pode ser descrita da seguinte maneira:

\begin{equation}
H = (r\cos\theta) I + \frac{1}{2}\omega \vec{\sigma}.\vec{\mu},
\end{equation}
onde ${\omega}$ representa a diferença de energia dos dois autoestados ($\epsilon_+ - \epsilon_-$) e $\vec{\mu}= \frac{2}{\omega}(s,0,ir\sin\theta)$.

Aplicando esta evolução temporal ao estado $|\nu_1 \rangle$, sucede que:

\begin{equation}
e^{\frac{-iH_{NH}t}{\hbar}}|\nu_1 \rangle = \frac{e^{\frac{-itrcos\theta}{\hbar}}}{cos\alpha} \left[\begin{array}{c}
cos(\frac{\omega t}{2\hbar}-\alpha)  \\
-isen(\frac{\omega t}{2\hbar})  \end{array} \right].
\end{equation}

Queremos calcular o tempo que o estado $|\nu_1 \rangle$ demora para evoluir até o estado $|\nu_2\rangle$ e, portanto, equacionamos:

\begin{equation} \left[\begin{array}{c}
cos(\frac{\omega\tau}{2\hbar}-\alpha)  \\
-isen(\frac{\omega\tau}{2\hbar})  \end{array} \right]
=
\left[\begin{array}{c}
0 \\
1 \end{array} \right]. 
\end{equation}

Deste modo, com $n$ $\in$ $\mathbb{N}$, o tempo de transição $\tau$ é dado por:

\begin{eqnarray}
 \frac{\omega\tau}{2\hbar} - \alpha = \frac{\pi}{2} + n.\pi, \\ 
\frac{\omega\tau}{2\hbar} = \alpha +  \frac{\pi}{2} + n.\pi,  \\ 
\tau = \frac{2\hbar(\alpha +  \frac{\pi}{2} + n.\pi)}{\omega}.
\end{eqnarray}

Note que o menor tempo possível é alcançado quando o natural $n$ é nulo e, portanto, o tempo otimizado para o processo é dado por $\tau^* = \tau(n=0)$:

\begin{equation}
\tau^* =  \frac{\hbar(2\alpha + \pi)}{\omega}.
\label{tau*}
\end{equation}

Se tomarmos o limite em que $\alpha \to - \pi/2$, na Eq. (\ref{tau*}), é possível obter um tempo de transição infinitesimalmente pequeno entre o estado $|\nu_1 \rangle$ e o estado $|\nu_2 \rangle$. 

Podemos ainda calcular a distância entre esses dois vetores, lembrando que o produto escalar entre dois estados normalizados é relacionado com o ângulo entre eles. Para calcular a distância angular entre $|\nu_1\rangle$ e $|\nu_2\rangle$, é necessário primeiramente normalizá-los com respeito a $\mathcal{C}PT$. Este procedimento é análogo ao realizado para os autoestados $|\epsilon_+\rangle$ e $|\epsilon_-\rangle$, conforme a condição de normalização (\ref{state1}).

Deste modo os estados $|v_1^{\prime}\rangle$ e $|v_2^{\prime}\rangle$ normalizados são:

\begin{eqnarray}
 |v_1^{\prime}\rangle = \sqrt{\cos \alpha}|v_1\rangle, \\
 |v_2^{\prime}\rangle = \sqrt{\cos \alpha}|v_2\rangle.
\end{eqnarray}

Com isso, seja $|u\rangle$ e $|v\rangle$, dois estados normalizados quaisquer, assim:

\begin{eqnarray}
 \langle u|v \rangle = \cos \beta, \\\
  \beta = \arccos{|\langle u|v \rangle_{\mathcal{C}PT}|},
 \label{distang}
\end{eqnarray}
onde $\beta$ é a distância angular entre os estados $|u\rangle$ e $|v\rangle$.

E, portanto, a distância angular entre os estados $|v_1^{\prime}\rangle$ e $|v_2^{\prime}\rangle$ é dada por:

\begin{equation}
\beta_{PT} = \arccos |\langle v_1\prime|v_2\prime \rangle_{\mathcal{C}PT}| = \arccos(|\sin\alpha|) .
\label{distnh}
\end{equation}

É interessante frisar que quando o tempo de transição vai a 0 ($\tau^* \rightarrow 0$), $\beta \rightarrow 0$. Ou seja, a distância entre os estados se torna pequena, produzindo vetores praticamente paralelos. Isto deve-se ao fato de que estes dois vetores não são mais ortogonais com respeito ao produto interno da Simetria PT, dependendo explicitamente do parâmetro $\alpha$ da Hamiltoniana.

Vamos também refazer esses cálculos para uma Hamiltoniana hermitiana, para efeitos de comparação. Seja $H_{H}$ uma Hamiltoniana geral Hermitiana, encontrada de maneira análoga a Hamiltoniana $H_{NH}$.

\begin{equation}
 H_{H} = \left[\begin{array}{c c}
s & r.e^{i\psi} \\
r.e^{-i\psi}  & u  \end{array} \right].
\label{hh}
\end{equation}

Nesta Hamiltoniana a diferença de energia entre seus dois estados ortonormais é dada por $\omega^{\prime}$

\begin{equation}
\omega^{\prime} = \sqrt{(s-u)^2 + 4r^2}.
\label{omegaprime}
\end{equation}

Expandindo a hamiltoniana em termos das matrizes de Pauli, exatamente como o realizado para a matriz $H_{NH}$, obtemos:

\begin{equation}
H_{H} = \frac{1}{2} (s + u) \vec{1} +\frac{1}{2} \omega^{\prime} \vec{\sigma}.\vec{n},
\label{expandh}
\end{equation}

Onde, $\vec{n}=\frac{2}{\omega^{\prime}}(r\cos\psi,-r\sin\psi,\frac{s-u}{2})$

Supondo que o estado inicial seja dado por $|\phi_i\rangle = (1,0)^t$ (Aqui t também se relaciona a operação transpor), a evolução do sistema pode ser descrita conforme a Eq. (\ref{sch}):

\begin{equation}
\exp[{-i H t/\hbar}] |\phi_i> = \left(\begin{array}{c}
a \\
b \end{array} \right),
\end{equation}

onde, uma vez que a evolução é unitária, $|a|^2+|b|^2 =1$. E então:

\begin{equation}
\left(\begin{array}{c}
a \\
b \end{array} \right)  = e^{-i(s+u)t/(2\hbar)} \left(\begin{array}{c}
\cos(\frac{\omega^{\prime} t}{2 \hbar}) -i \frac{s-u}{\omega^{\prime}}\sin(\frac{\omega^{\prime} t}{2\hbar}) \\
-i\frac{2r}{\omega^{\prime}} \sin(\frac{\omega^{\prime} t}{2 \hbar})e^{-i\psi}\end{array} \right).
\end{equation}

Analisando a segunda linha das matrizes é possível encontrar a relação:

\begin{equation}
|b| = \frac{2r}{\omega^{\prime}}\sin\left(\frac{\omega^{\prime} t}{2 \hbar}\right) .
\end{equation}

E portanto o tempo de transição é descrito por:

\begin{equation}
 t = \frac{2 \hbar}{\omega^{\prime}}\arcsin\left(\frac{\omega^{\prime} |b|}{2 r}\right).
 \label{fleeming2}
\end{equation}

Para otimizar essa relação podemos encontrar o valor máximo de r em função de $\omega^{\prime}$, a Eq. (\ref{omegaprime}) nos diz que isso ocorre quando $s=u$ e, portanto, $r= \frac{\omega^{\prime}}{2}$.  E com isso podemos escrever:

\begin{equation}
\tau = \frac{2 \hbar}{\omega^{\prime}}\arcsin\left(|b|\right).
\label{fleeming3}
\end{equation}

E então, como conclusão, se comparamos um estado $|\phi_i\rangle = (1,0)^t$ com um estado $|\phi_f\rangle = (a,b)^t = (0,1)^t$ (lembre-se que $|a|^2+|b|^2 = 1$ e se $b \rightarrow 0$, $a \rightarrow 1$), vemos que o tempo de transição entre 2 estados ortogonais não pode ser maior que:

\begin{equation}
 t_f = \frac{\hbar \pi}{\omega}.
 \label{fleming}
 \end{equation}

Note que o tempo de transição depende explicitamente do estado final, Eq. (\ref{fleeming3}). Escolhendo o estado final como $|\phi_f \rangle \rightarrow (1,0)^t$, temos que $b \rightarrow 0$, e segundo a Eq. (\ref{fleeming2}), segue que $t\rightarrow 0$.

 Assim como o realizado para o caso não Hermitiano, podemos calcular a distância entre os estados, $|\phi_i\rangle = (1,0)^t$ e $|\phi_f\rangle = (a,b)^t$:

\begin{equation}
\beta_h = \arccos(|<\phi_i|\phi_f>|) = \arccos{(|a|)}.
\label{b}
\end{equation}

E então, quando calculamos o tempo de transição entre estados para o caso da Mecânica Quântica PT Simétrica (\ref{hnm}), notamos que quando o parâmetro $\alpha$ tendia a $-\pi/2$ ($\alpha \rightarrow -\pi/2$), obtínhamos um tempo de transição $t$ tendendo a 0 ($t \rightarrow 0$), enquanto que, quando o parâmetro $\alpha$ tendia a 0 ($\alpha \rightarrow 0$), obtínhamos um tempo de transição $t$ tendendo a $\hbar \pi/\omega$ ($t \rightarrow \hbar \pi/\omega$). Para o caso da Hamiltoniana hermtiana, (\ref{hh}), quando nosso parâmetro $b$ vai a 0 ($b \rightarrow 0$), o tempo de transição $t$ também vai a 0 ($t \rightarrow 0$), no entanto, quando o parâmetro $b$ tende a 1 ($b \rightarrow 1$), o tempo de transição $t$ tende a $\hbar \pi/\omega^{\prime}$ ($t \rightarrow \hbar \pi/\omega^{\prime}$). E através disso podemos entender o que está acontecendo com os estados, no caso de Simetria PT, quando $\alpha \rightarrow -\pi/2$, os estados são praticamente degenerados, uma vez que:

\begin{eqnarray}
 |v_1^{\prime}\rangle = \sqrt{\cos \alpha}|v_1\rangle, \\
 |v_2^{\prime}\rangle = \sqrt{\cos \alpha}|v_2\rangle,
\end{eqnarray}
$ |v_1^{\prime}\rangle \approx |v_2^{\prime}\rangle$ e por isso $\beta_{PT} \rightarrow 0$ e, portanto, esperaríamos poder fazer uma transição tão rápida quanto desejada. Quando $\alpha \rightarrow 0$, temos justamente o caso onde a hamiltoniana (\ref{hnm}) se torna hermitiana, uma vez que $0 = \sin\alpha = (r/s)\sin\psi \Rightarrow \psi = 0$, como consequência, o produto interno $\mathcal{C}PT$ se torna o produto interno usual de Dirac, assim, $|v_1^{\prime}\rangle$ se torna ortogonal a $|v_2^{\prime}\rangle$ e por isso $\beta_{PT} \rightarrow \pi/2$.
Analogamente, para o caso da hamiltoniana (\ref{hh}), quando o parâmetro $b \rightarrow 0$, estamos fazendo a evolução entre dois estados quase degenerados $|\phi_i> = (1,0)^t$ e $|\phi_f> = (a,b)^t$ e por isso $\beta_h \rightarrow 0$, todavia, quando $b \rightarrow 1$, temos a transição entre dois estados ortogonais e por isso $\beta_h \rightarrow \pi/2$. Ou seja, o efeito é o mesmo nos dois casos, apenas há uma má interpretação dos estados no artigo \cite{faster} e, portanto, não existe nenhum comportamento de transição mais rápida entre estados em Mecânica Quântica PT Simétrica em contraste à Mecânica Quantica Usual.

 \section{Conclusão}
 
Ao analisar as condições de Simetria PT pode-se mostrar a equivalência entre a Mecânica Quântica Usual e a Mecânica Quântica PT Simétrica. Aplicar as bases deste novo formalismo não fornece novidade alguma  ao sistema, em contraste ao que é sugerido por \cite{faster}. Sem perda de generalidade pode-se descrever todo sistema quântico com matrizes Hermitianas. Uma demonstração matematicamente formal desta afirmação pode ser encontrada em \cite{mustafa}.  Nesse artigo concluímos que o efeito de transição entre estados quânticos de maneira mais rápida que Mecânica Quântica Usual apenas se trata de uma interpretação equivocada do comportamento dos estados. Quando analisamos o comportamento físico dos estados através de suas distâncias angulares, percebemos que o efeito de transição rápida ocorre tanto em Mecânica Quântica PT Simétrica quanto em Mecânica Quântica Usual, no entanto, essa transição infinitamente rápida somente ocorre entre estados quase paralelos e que, tanto em Simetria PT, quanto em Mecânica Quântica Usual o tempo de transição entre estados ortogonais é limitada por um tempo mínimo de $t = \hbar \pi/\omega$, sendo $\omega$ a diferença de energia dos dois estados ortogonais.

\bibliographystyle{science} 
\bibliography{references} 

\end{document}